\documentclass[a4paper]{article}

\usepackage{multirow}
\usepackage{float}
\usepackage{amsfonts}

\usepackage{INTERSPEECH2020}

\title{DNN Speaker Tracking with Embeddings}
\name{Carlos Rodrigo Castillo-Sanchez$^1$, Leibny Paola Garcia-Perera$^2$, Anabel Martin-Gonzalez$^1$}
\address{
  $^1$Computational Learning and Imaging Research, Universidad Autónoma de Yucatán, Mexico\\
  $^2$Center for Language and Speech Processing, Johns Hopkins University, USA}
\email{carloscastillomvc@gmail.com, leibny@gmail.com, amarting@correo.uady.mx}

\begin{document}

\maketitle
\begin{abstract}
In multi-speaker applications is common to have pre-computed models from enrolled speakers. Using these models to identify the instances in which these speakers intervene in a recording is the task of speaker tracking.  
  In this paper, we propose a novel embedding-based speaker tracking method. Specifically, our design is based on a convolutional neural
  network that mimics a typical speaker verification PLDA (probabilistic linear discriminant analysis) classifier and finds the regions uttered by the target speakers in an
  online fashion. The system was studied from two different perspectives: diarization and tracking; results on both
 show a significant improvement over
  the PLDA baseline under the same experimental conditions.
  Two standard public datasets, CALLHOME and DIHARD II single
  channel, were modified to create
  two-speaker subsets with overlapping and non-overlapping regions. 
  We evaluate the robustness of our supervised approach with models generated from different segment lengths.
  A relative improvement of 17\% in DER for DIHARD II single channel shows promising performance.
  Furthermore, to make the  baseline system similar to a speaker tracking, non-target speakers were added to the recordings. Even in these adverse conditions, our approach is robust enough to outperform the PLDA baseline.
\end{abstract}

\section{Introduction}
  Speaker tracking is the process of identifying all regions uttered by a target speaker in an audio
  ~\cite{ASTSBOSTDFNE}. Similarly to speaker diarization, which  answers the question \textit{"who spoke when?"}, speaker tracking searches for those regions, but assigns speaker identities. 
  This process is an important pre-processing step for many multi-speaker applications such as virtual
  assistants and broadcast news transcription and indexing~\cite{8272968}.

As shown in \cite{garcia2019speaker}, diarization and tracking are two methods closely related. Although tracking would benefit from the diarization, in this research, we explored the possibility to include a neural network as a robust classifier that can operate similarly to the PLDA. The goal is that it can naturally provide results for diarization and tracking. Since there are just a few studies on speaker tracking \cite{ASTSBOSTDFNE,8272968,sonmez1999speaker}, we use diarization as the main background and inspiration of this work.

  Most of the standard speaker diarization systems focus on offline clustering as it uses all the
  contextual information to label the speech regions. Examples of such algorithms include agglomerative
  hierarchical clustering (AHC)~\cite{article:bsdfdsdc2019,article:sdwdsefdci}, k-means~\cite{article:pal2019study,
  article:hdwkfscisdobn}, spectral clustering~\cite{article:lbsmwscfsd,article:atscfsdunme}, etc. 
  These clustering methods cannot be used in real-time applications since they require the complete speech data
  upfront. If the application is latency-sensitive it requires to have speaker labels generated as soon as speech
  segments are available for the system.

  ~\cite{wang2017speaker} presents an embedding based speaker diarization system. \textit{d-vectors} were used
  ~\cite{article:dnnfsftdsv} along with an LSTM-based speaker verification in combination with spectral clustering to
  successfully perform offline diarization; however, the diarization error rate almost doubles in its online modality.
  Another online diarization approach is introduced in ~\cite{Ghahabi2019}. They propose a DNN (deep neural network) embedding suitable for
  online processing referred as speaker-corrupted embedding. The diarization algorithm uses cosine similarity to
  compare the speaker models and the embedding in order to make the labeling decisions.

  \begin{figure}
    \centering
    \includegraphics[width=\linewidth]{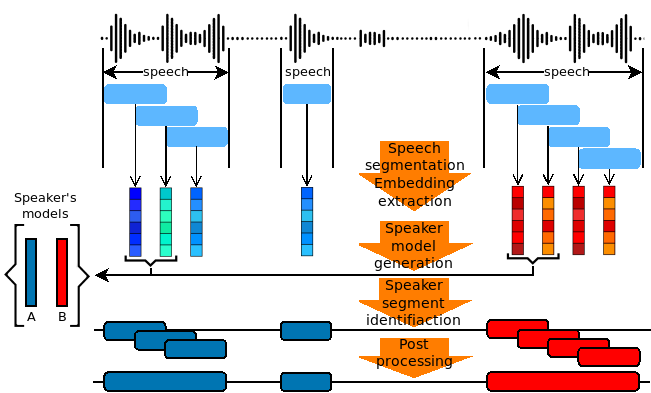}
    \caption{Pipeline of proposed speaker tracking system.}
    \label{fig:pipeline}
  \end{figure}

  In this paper, we propose an online speaker tracking pipeline by replacing the
  unsupervised offline clustering module from the standard diarization system with a online tracking method that uses
  a DNN as an embedding robust classifier.
  As shown in Figure~\ref{fig:pipeline}, our speaker tracking system shares many of its components with the standard
  diarization pipeline~\cite{article:bsfdsdc2018,zhang2018fully,dihseallftjtitidc}, with the main difference being the
  clustering algorithm.

  The experimental results on CALLHOME and DIHARD II single channel reveal that our method achieves significant
  improvements over the PLDA baseline.
  \footnote{ Our code is available at https://github.com/CarlosRCS9/kaldi/tree/paper-dnn-tracking}

\section{Methodology}
  In this section, we introduce our speaker tracking framework, Figure~\ref{fig:pipeline} illustrates the overall steps of our tracking pipeline.

  \begin{itemize}
    \item Speech segmentation and embedding extraction.
    \item Speaker model generation.
    \item Speaker segment identification.
    \item Post-processing.
  \end{itemize}

  \subsection{Speech segmentation and embedding extraction}
    The first module in our pipeline is inspired by the standard diarization system, it uses a Voice Activity Detector (VAD) to determine the speech regions in the input audio signal, excluding
    the non-speech regions from subsequent processing. A sliding window, further divides these regions into a set of
    smaller, overlapping speech segments, establishing the temporal resolution of the speaker tracking results. The output of this module is a set of voiced-speech segments. 
    We decided to use an oracle VAD as segmentation mechanism to focus our efforts on checking whether our
    proposed architecture can track speakers accurately.

    \subsubsection{Embedding extraction}
      The next step in the pipeline is to extract an embedding from each segment. Our system was
      tested following the i-vector- and x-vector-based approaches~\cite{article:sdwpisauc,article:xrdefsr}.
      The i-vector, introduced by Dehak \textit{et al.}~\cite{article:fefafsv}, is a speaker representation that
      provides a way to reduce large-dimensional input speech data to a small-dimensional feature vector that retains
      most of the relevant channel and speaker information. 
      The x-vector, introduced by Snyder \textit{et al.}~\cite{article:dnnbsefetesv,article:xrdefsr} is an embedding
      extracted from a deep neural network trained to discriminate between speakers, mapping variable-length
      speech segments to a fixed-length feature vector. 
      Nowadays, the
      x-vector approach provides state-of-the-art performance in many speaker recognition fields, such as, speaker
      verification and speaker diarization~\cite{article:srfmscux,article:slrux,article:gxftixv,ryant2019second}.

  \subsection{Speaker model generation}
    After the segment embeddings are extracted, a speaker model is generated for each tracked speaker. Such task is
    performed by averaging the embeddings within a time window at the beginning of each target speaker enrollment in the
    input audio. We define \textit{model time} as the window width used to generate the speaker's models. With this
    approach, the system operates in an online fashion in which with a few labeled samples of the target speakers it can
    find their appearances along the complete audio. 

  \subsection{Speaker segment identification}
    The resulting segment embeddings and the speaker models are then passed through a speaker identification/verification stage.
    This task is performed by a speaker tracking
    DNN, the key component of our pipeline. According to the run-time latency, this module follows an \textbf{online}
    tracking strategy. It produces a speaker label immediately after a segment is available without the knowledge of future
    segments, making it easier for the system to deal with large amounts of data.
    
    \subsubsection{Features}
    
      \begin{figure}
        \centering
        \includegraphics[width=\linewidth]{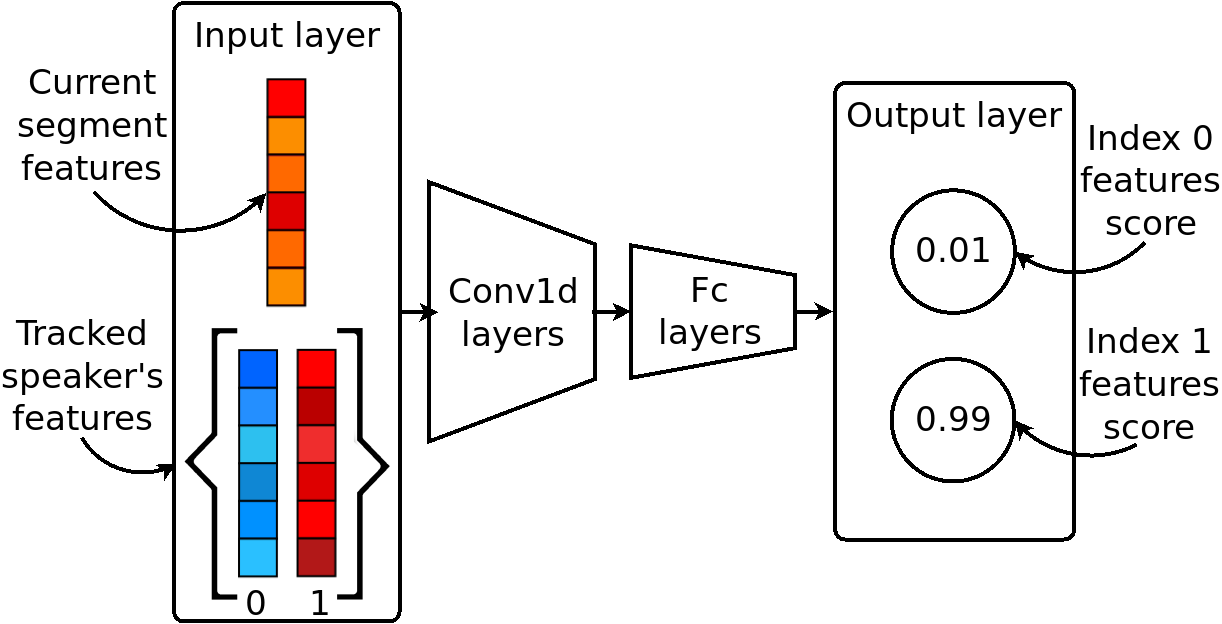}
        \caption{Network input and output layers during the identification process.}
        \label{fig:architecture}
      \end{figure}
    
      Figure~\ref{fig:architecture} illustrates the structure of the network's input and output layers during the
      segment labeling process. For a given utterance the input and output sequences of the network ($X'$, $Y'$) are
      defined as follows:

      \begin{itemize}
        \setlength\itemsep{0em}
        \item The speech segmentation and embedding extraction module provides a sequence of embeddings
        $X=(x_1, x_2, \ldots, x_T)$, where each $x_t\in\mathbb{R}^b$ has a 1:1 correspondence to
        the $T$ segments obtained from the input utterance, and $b$ is the dimension of every embedding.
        \item The speaker model generation module provides the sequence $M=(m_1, m_2, \ldots, m_S)$ where
        $m_s\in\mathbb{R}^b$,  such that each entry of the sequence is a model of one of the $S$ tracked speakers.
        \item The input sequence of our network is defined as the concatenation of $M$ to each element of $X$.
        $X'=\{x_t^\frown M|x_t\in X\}$.
        \item The sequence $Y=(y_1, y_2, \ldots, y_T)$ is given by the speaker labels of the $T$
        segments.
        \item The output sequence is given by $Y'=\{\Phi(y_t)|y_t\in Y\}$ where $\Phi(y_t)=\{P(m_s|x_t,y_t)|m_s\in M\}$. At training time $Y$ is given by the ground-truth labels. At inference, $Y$ is computed by the estimated labels. 
      \end{itemize}

    \subsubsection{Architecture}
    
     Table~\ref{tab:architecture} summarizes the final DNN architecture used in this work. The first three convolutional layers of the network provide a comparison stream for each of the $S$ speakers models, where the similarity measure between the segment embedding and a single speaker model is computed with the contextual information of all the speaker models. Then the fully-connected feed-forward layers use these streams to score the similarity of the target speaker model and the incoming segment. \footnote {We also tested recurrent neural network architectures (bidirectional LSTM), but 
      it was discarded as it seemed to ignore the speaker model features at inference time, and remembered the input sequence established at training phase.}
     
      \begin{table}[t]
        \centering
        \tabcolsep=0.11cm
        \begin{tabular}{lccc}
          \toprule
          \textbf{Layer type} & \textbf{\# filters}    & \textbf{Kernel size}   & \textbf{Input $\times$ output}   \\
          \midrule
          Conv1d.ReLU         & $S^3$                  & $3$                    & $b(S+1) \times (b-2)S^3$         \\
          Conv1d.ReLU         & $S^2$                  & $3$                    & $(b-2)S^3 \times (b-4)S^2$       \\
          Conv1d.ReLU         & $S$                    & $3$                    & $(b-4)S^2 \times (b-6)S$         \\
          Dense.ReLU          &                        &                        & $(b-6)S \times 32S$              \\
          Dense.ReLU          &                        &                        & $32S \times 16S$                 \\
          Dense.Sigmoid       &                        &                        & $16S \times S$                   \\
          \bottomrule
        \end{tabular}
        \caption{DNN speaker tracking architecture.}
        \label{tab:architecture}
        \vspace{-1em}
      \end{table}

    \subsubsection{Training}
\begin{figure}[H]
        \centering
        \includegraphics[width=\linewidth]{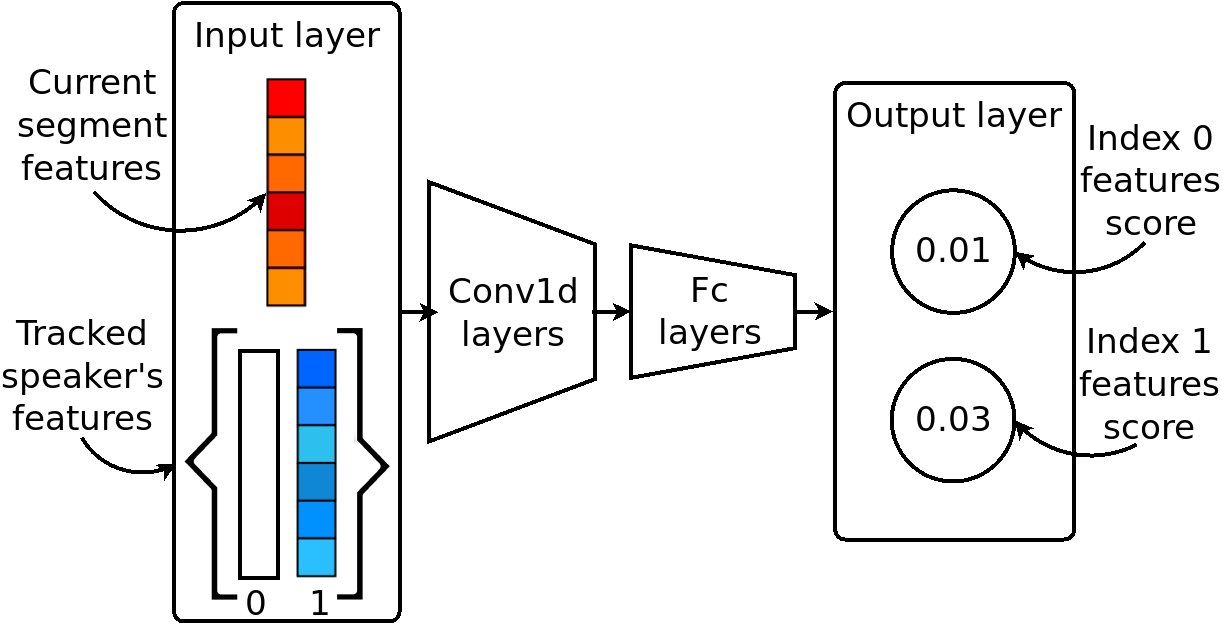}
        \caption{Input layer with a non-speaker embedding during the network training.}
        \label{fig:architecture2}
      \end{figure}

      During training, all possible permutations of the elements of $M$ are computed and appended to every input $x_t$ with two main goals:
      reduce overfitting by forcing all output neurons to score the same speaker models, and augment the number of
      training samples. This procedure ensures the DNN scoring to be independent of the speaker model sequence order.
      Figure~\ref{fig:architecture2} shows how the training data is furthermore augmented with the addition of zero
      padding as non-speaker model feature. This procedure simulates a 
      verification task since the network has to decide whether the current segment embedding belongs to one of the available 
      models or not.
      
      At inference time, the input layer of the network receives the incoming segment embedding and an array of target speakers models, the length of the array is the same as output neurons, so each score is related
      to an index in the speakers models array. In an identification setup we label the segment with the highest score index. If the task requires verification, a certainty threshold is used to label the segments.
      
  \subsection{PLDA}
    The baseline system uses probabilistic linear discriminant analysis (PLDA) scoring as the similarity measure\footnote{PLDA scoring computes the loglikehood ratio between two embeddings}, since it has proven to achieve state-of-the-art performance in many speaker
    recognition tasks. It provides a powerful distortion-resistant mechanism to distinguish between different speakers, and robust to same speaker variability.
    ~\cite{article:apaflatisr,article:ltfpisv,ryant2019second,article:unsdsftdi2c}.

  \subsection{Post-processing}
    Due to the \textbf{online} nature of our pipeline, the post-processing step is applied as soon as a segment label is
    emitted, this step refines the tracking results by merging the same-speaker contiguous segments. And also 
    by adjusting the labels within a window of three contiguous segments $W_t=(x_{t-2},x_{t-1},x_{t})$, it modifies the
    in-between segment label if the surrounding labels are equal to each other and differ from the in-between label,
    producing three contiguous segments with the same label.

\section{Experiments}
  This section describes our experimental setup and results. We decided on a 1 s width and 0.5 s step sliding window at
  the speech segmentation step, discarding segments shorter than 0.5 s to ensure sufficient speaker information.
  Both i- and x-vectors were extracted using the Kaldi's CALLHOME diarization recipes\footnote{https://github.com/kaldi-asr/kaldi/tree/master/egs/callhome diarization/v1 and /v2}
  ~\cite{dihseallftjtitidc}. For CALLHOME x-vector experiments, a publicly available\footnote{https://kaldi-asr.org/models/m6}~\cite{article:xrdefsr} model and PLDA backend were used.

  \subsection{Evaluation metrics}
    The system performance was evaluated in terms of Equal Error Rate (EER) and minimum Detection Cost Function (minDCF),
    as the key component of our tracking framework follows a speaker verification approach. In addition, we report
    Diarization Error Rate (DER)~\cite{article:trt2smre} since our framework shares characteristics with the
    standard diarization system.

  \subsection{Datasets}
    We tested our system on two standard public datasets: (1) 2000 NIST Speaker Recognition Evaluation (LDC2001S97),
    Disk-8, usually referred to as “CALLHOME”, it contains 500 utterances distributed across six languages: Arabic,
    English, German, Japanese, Mandarin, and Spanish. Each utterance contains up to 7 speakers; (2) DIHARD II single
    channel development and evaluation subsets (LDC2019E31, LDC2019E32), focused on "hard" speaker diarization, contains
    5-10 minute English utterances selected from 11 conversational domains, each including approximately 2 hours of
    audio. Since our approach is supervised, we perform a 2-fold cross-validation on each dataset using standard
    partitions: callhome1 and callhome2 from Kaldi's CALLHOME diarization recipe~\cite{dihseallftjtitidc}, and DIHARD II
    single channel's development and evaluation subsets. Then, the partitions results are combined to report the averaged
    DER, ERR and minDCF of each dataset.
    
  \subsection{Overlap preparation}
    \label{sec:overlap}
    A set of our experiments is focused on speaker overlap, so it was necessary to augment the datasets, as they have a low percentage of speaker overlap (CALLHOME $\sim$16\%,
    DIHARD II single channel $\sim$9\%). To perform this task the non-overlapping audio segments of each speaker
    are extracted using the ground-truth labels, then merged into a set of single-speaker utterances for each recording. After that, the single-speaker
    utterances are pairwise overlapped to create a new set of two-speaker-overlapping utterances. Finally,
    the new overlapping utterances are cut into segments and inserted into their original recordings at random
    locations. The resulting datataset contains an additional $\sim$18\% of speaker overlap in CALLHOME, and $\sim$30\% in DIHARD
    II single channel.
  
  \subsection{Baseline}
    The baseline system follows exactly the same procedure as our proposed tracking method. The only difference is
    the replacement of the DNN-based speaker segment identification module with a PLDA-based one.

  \subsection{Results}
    In the first set of experiments we provide an optimum set of conditions for speaker tracking, the number of tracked
    speakers is fixed to 2, with the input audio signal containing only speech from them, and there is no overlapped
    speech instances.
    
    \begin{table}[t]
      \centering
      \tabcolsep=0.11cm
      \caption{DER (\%), EER (\%) and minDCF (0.1\% target probability) on two datasets given the optimum conditions.}
      \label{tab:optimum}
      \begin{tabular}{lcccc}
        \toprule
        \multirow{2}{*}{\textbf{Model time}} & \multicolumn{2}{c}{\textbf{PLDA}} & \multicolumn{2}{c}{\textbf{DNN}} \\
        \cmidrule(r{4pt}){2-3} \cmidrule(l){4-5}
        & \multicolumn{1}{c}{\textbf{DER}} & \multicolumn{1}{c}{\textbf{EER \tiny{minDCF}}} & \multicolumn{1}{c}{\textbf{DER}} & \multicolumn{1}{c}{\textbf{EER \tiny{minDCF}}} \\
        \midrule
        \multicolumn{5}{l}{\textbf{CALLHOME i-vector}}                                                                                                                          \\
        3.0 s  & 6.90                      & 17.40 \tiny{0.91}                              & 5.65                             & 4.53 \tiny{0.52}                               \\
        5.5 s  & 5.41                      & 14.81 \tiny{0.88}                              & 4.93                             & 3.90 \tiny{0.53}                               \\
        10.5 s & 4.56                      & 12.98 \tiny{0.85}                              & 4.54                             & \textbf{2.83} \tiny{\textbf{0.37}}             \\
        \multicolumn{5}{l}{\textbf{x-vector}}                                                                                                                                   \\ 
        3.0 s  & 33.95                     & 41.66 \tiny{0.99}                              & 11.53                            & 11.23 \tiny{0.79}                              \\
        5.5 s  & 28.65                     & 36.68 \tiny{1.00}                              & 7.80                             & 6.93 \tiny{0.60}                               \\
        10.5 s & 24.66                     & 32.57 \tiny{1.00}                              & \textbf{5.63}                    & 4.17 \tiny{0.50}                               \\
        \midrule
        \multicolumn{5}{l}{\textbf{DIHARD II i-vector}}                                                                                                                         \\
        3.0 s  & 19.01                     & 36.62 \tiny{0.99}                              & 18.44                            & 21.53 \tiny{0.96}                              \\
        5.5 s  & 16.22                     & 34.73 \tiny{0.99}                              & 13.97                            & 15.19 \tiny{0.89}                              \\
        10.5 s & 13.29                     & 33.70 \tiny{0.99}                              & 11.03                            & \textbf{11.76} \tiny{\textbf{0.82}}            \\
        \multicolumn{5}{l}{\textbf{x-vector}}                                                                                                                                   \\
        3.0 s  & 28.49                     & 38.01 \tiny{1.00}                              & 20.56                            & 29.01 \tiny{0.99}                              \\
        5.5 s  & 28.54                     & 38.01 \tiny{1.00}                              & 19.28                            & 27.04 \tiny{0.99}                              \\
        10.5 s & 27.05                     & 39.69 \tiny{1.00}                              & \textbf{13.36}                   & 17.51 \tiny{0.97}                              \\
        \bottomrule
      \end{tabular}
      \vspace{-1em}
    \end{table}

\begin{figure}[ht]
      \centering
      \includegraphics[width=\linewidth]{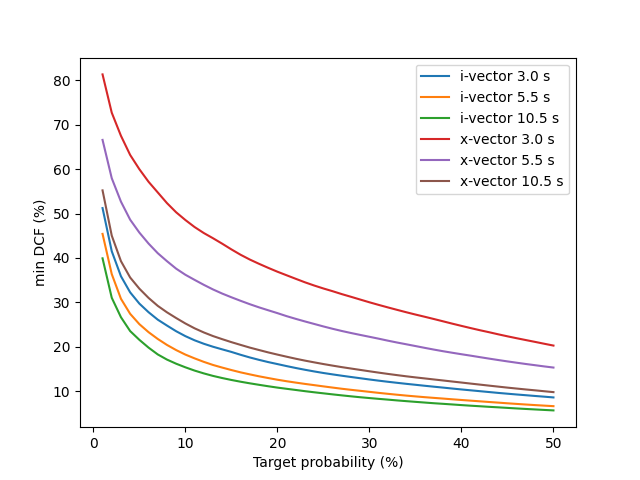}
      \caption{Model time minDCF curves for i- and x-vector with DNN similarity measure, given the optimum conditions.}
      \label{fig:mindcf_target_probability}
    \end{figure}

    In Table~\ref{tab:optimum} we can see that the DNN based tracking system significantly outperforms the PLDA
    baseline in EER and minDCF, which drops from 12.98\% to 2.83\% and 0.85 to 0.37, respectively in CALLHOME with the
    speaker models generated with 10.5 s of labeled samples. We also observe that the i-vectors provide good DER
    performance even when fewer labeled samples are provided. Additionally, we followed the same optimum conditions with x-vectors, the advantage of our
    supervised approach in EER and minDCF is consistent in both datasets.
    A further improvement is shown in terms of DER, which drops form 24.66\% to 5.63\%. It should be noted that
    both, DNN and PLDA systems, degrade when x-vectors are used, this is further illustrated in
    Figure~\ref{fig:mindcf_target_probability}, where the minDCF curves show a clear advantage of the i-vectors in our
    supervised approach. We attribute this behavior to the training of the x-vectors (1.5 seconds window) and the very short segments used to produce the embeddings.

    \begin{table}[t]
      \centering
      \tabcolsep=0.11cm
      \caption{DER (\%), EER (\%) and minDCF (0.1\% target probability) given the speaker verification conditions.}
      \label{tab:verification}
      \begin{tabular}{lcccc}
        \toprule
        \multirow{2}{*}{\textbf{Model time}} & \multicolumn{2}{c}{\textbf{PLDA}} & \multicolumn{2}{c}{\textbf{DNN}} \\
        \cmidrule(r{4pt}){2-3} \cmidrule(l){4-5}
        & \multicolumn{1}{c}{\textbf{DER}} & \multicolumn{1}{c}{\textbf{EER \tiny{minDCF}}} & \multicolumn{1}{c}{\textbf{DER}} & \multicolumn{1}{c}{\textbf{EER \tiny{minDCF}}} \\
        \midrule
        \multicolumn{5}{l}{\textbf{CALLHOME i-vector}}                                                                                                                          \\
        3.0 s  & 7.37                      & 22.58 \tiny{0.99}                              & 6.25                             & 11.66 \tiny{0.79}                              \\
        5.5 s  & 5.69                      & 22.52 \tiny{0.99}                              & 4.56                             & 7.77 \tiny{0.64}                               \\
        10.5 s & 4.53                      & 22.39 \tiny{1.00}                              & 4.43                             & \textbf{5.89} \tiny{\textbf{0.52}}             \\
        \multicolumn{5}{l}{\textbf{x-vector}}                                                                                                                                   \\ 
        3.0 s  & 32.57                     & 46.58 \tiny{1.00}                              & 12.37                            & 6.81 \tiny{0.67}                               \\
        5.5 s  & 27.61                     & 41.52 \tiny{1.00}                              & 8.13                             & 5.26 \tiny{0.66}                               \\
        10.5 s & 22.81                     & 38.06 \tiny{1.00}                              & 6.24                             & \textbf{4.40} \tiny{\textbf{0.65}}             \\
        \midrule
        \multicolumn{5}{l}{\textbf{DIHARD II i-vector}}                                                                                                                         \\
        5.5 s  & 16.40                     & 32.86 \tiny{0.99}                              & 15.42                            & 17.96 \tiny{0.94}                              \\
        10.5 s & 13.05                     & 32.67 \tiny{0.99}                              & 11.44                            & \textbf{14.67} \tiny{\textbf{0.96}}            \\
        \multicolumn{5}{l}{\textbf{x-vector}}                                                                                                                                   \\
        5.5 s  & 29.48                     & 56.65 \tiny{1.00}                              & 16.84                            & 15.89 \tiny{0.98}                              \\
        10.5 s & 30.04                     & 59.99 \tiny{1.00}                              & 13.77                            & \textbf{15.06} \tiny{\textbf{0.99}}            \\
        \bottomrule
      \end{tabular}
      \vspace{-1em}
    \end{table}

    In a second set of experiments we increased the complexity of the previous conditions by mixing a set of
    non-target speaker segments in every recording. Such segments were built from the speaker models of other
    recordings within the same cross-validation fold. We are interested in the EER and minDCF results.

    As shown in Table~\ref{tab:verification} the DNN-based
    system continues to outperform the PLDA baseline. We also observe that the performance gap between i- and x-vector
    disappears in the DNN system, since the i-vector error increased while the x-vector's pretty much remained the same. We can conclude that our system is robust enough when it encounters non-target speakers along the recording.

    Finally, we evaluate our proposed system considering overlapped speech, as described in Section \ref{sec:overlap}. In this set of experiments, the number of tracked
    speakers is fixed to 2, with the input audio signal containing non-overlapping and overlapping speech from them,
    Table~\ref{tab:overlap} shows promising results in both datasets, with 12\% DER in CALLHOME with the additional 18\% of augmented overlapped speech; and 28.04\% DER in DIHARD II single channel with its 30\% additional overlap.

    \begin{table}[t]
      \centering
      \tabcolsep=0.11cm
      \caption{DER (\%), EER (\%) and minDCF (0.1\% target probability) for \textbf{i-vector}, given the speaker overlap conditions.}
      \label{tab:overlap}
      \begin{tabular}{lcccc}
        \toprule
        \multirow{2}{*}{\textbf{Model time}} & \multicolumn{2}{c}{\textbf{CALLHOME}} & \multicolumn{2}{c}{\textbf{DIHARD II}} \\
        \cmidrule(r{4pt}){2-3} \cmidrule(l){4-5}
        & \multicolumn{1}{c}{\textbf{DER}} & \multicolumn{1}{c}{\textbf{EER \tiny{minDCF}}} & \multicolumn{1}{c}{\textbf{DER}} & \multicolumn{1}{c}{\textbf{EER \tiny{minDCF}}} \\
        \midrule                                                                                                                                                            
        3.0 s  & 20.82                     & 13.20 \tiny{0.94}                              & 36.94                            & 30.35 \tiny{0.99}                              \\
        5.5 s  & 15.78                     & 9.72 \tiny{0.90}                               & 31.99                            & 24.78 \tiny{0.98}                              \\
        10.5 s & \textbf{12.64}            & 7.55 \tiny{0.83}                               & \textbf{28.04}                   & 20.63 \tiny{0.94}                              \\
        \bottomrule
      \end{tabular}
      \vspace{-1em}
    \end{table}

\section{Conclusions}
  In this paper, we propose a novel embedding-based speaker tracking DNN model focused on online tracking. We
  demonstrated efficiency of our approach through several experiments on two standard public datasets: CALLHOME and DIHARD II
  single channel. Validation results show a promising performance improvement compared to the PLDA baseline, as it drops the DER, EER
  and minDCF in the different experimental conditions, such as increased number of non-target speakers within a recording, and overlapping speakers. 

  For future research, we would like to extend our current DNN model to an \textbf{online} diarization and tracking system, where a recurrent
  neural network (RNN) will be responsible for selecting and updating the speaker models without having to resort to external sources. We expect such system to provide not only the diarization results, but also the set of speaker
  models that it will generate during an adaptive diarization process.

\section{Acknowledgments}
  We would like to thank Diego Nigel Joaquin Campos Sobrino and Mario Alejandro Campos Soberanis for their helpful discussions.

\bibliographystyle{IEEEtran}

\bibliography{mybib}
\end{document}